\documentclass{icm2010}
\usepackage{graphicx}

\title[{Discrete Complex Analysis and Probability}]
{Discrete Complex Analysis\\ and Probability}
\author[Stanislav Smirnov]{Stanislav Smirnov
\thanks{This research was supported by the European Research Council AG ``CONFRA'' 
and by the Swiss National Science Foundation.
We would like to thank Dmitry Chelkak for comments on the preliminary version of this paper.}
}
\contact[stanislav.smirnov@unige.ch]{
Section de Math\'ematiques, Universit\'e de Gen\`eve.
2-4 rue du Li\`evre, CP 64, 1211 Gen\`eve 4, SUISSE}

\newtheorem{theorem}{Theorem}

\newtheorem{lemma}[theorem]{Lemma}
\newtheorem{cor}{Corollary}
\theoremstyle{definition}

\newtheorem{remark}{Remark}
\newtheorem{question}{Question}
\def\br#1{\left(#1\right)}
\def\brl#1{\left\langle#1\right\rangle}
\def\brs#1{\left\{#1\right\}}
\def\abs#1{\left|#1\right|}
\def\spin{s}
\def\black{}
\def\red{}
\def\blue{}
\newcommand{\PP}{{\mathbb P}}

\newcommand{\ZZ}{{\mathbb Z}}
\newcommand{\RR}{{\mathbb R}}
\newcommand{\CC}{{\mathbb C}}

\newcommand{\IM}{{\mathrm{Im}}}

\newcommand{\unif}{\rightrightarrows}

\begin{document}

\begin{abstract}
We discuss possible discretizations of complex analysis
and some of their applications to probability
and mathematical physics, following our recent work
with Dmitry Chelkak, Hugo Duminil-Copin and Cl\'ement Hongler.
\end{abstract}



\maketitle

\section{Introduction}\label{sec:intro}

The goal of this note is to discuss some of the
applications of discrete complex analysis
to problems in probability and statistical physics.
It is not an exhaustive survey, and it lacks many references.
Forgoing completeness, we try to give
a taste of the subject through examples,
concentrating on a few of our recent papers
with Dmitry Chelkak, Hugo Duminil-Copin and Cl\'ement Hongler 
\cite{chelkak-smirnov-dca,chelkak-smirnov-iso,
chelkak-smirnov-spin,duminil-smirnov-saw,hongler-smirnov-density}.
There are certainly other interesting
developments in discrete complex analysis,
and it would be a worthy goal to write an extensive exposition
with an all-encompassing bibliography,
which we do not attempt here for lack of space.

Complex analysis 
(we restrict ourselves to the case of one complex or equivalently two real dimensions) 
studies analytic functions on (subdomains of) the complex plane,
or more generally analytic structures on two dimensional manifolds.
Several things are special about the (real) dimension two,
and we won't discuss an interesting and often debated question,
why exactly complex analysis is so nice and elegant.
In particular, several definitions lead to identical class of analytic functions,
and historically different adjectives (regular, analytic, holomorphic, monogenic) 
were used, depending on the context.
For example, an \emph{analytic} function has a local power series expansion around every point,
while a \emph{holomorphic} function has a complex derivative at every point.
Equivalence of these definitions is a major theorem in complex analysis,
and there are many other equivalent definitions in terms of Cauchy-Riemann equations, 
contour integrals, primitive functions, 
hydrodynamic interpretation, etc.
Holomorphic functions have many nice properties,
and hundreds of books were devoted to their study.

Consider now a discretized version of the complex plane:
some graph embedded into it, say a 
square or triangular lattice
(more generally one  can speak of discretizations of Riemann surfaces).
Can one define analytic functions on such a graph?
Some of the definitions do not admit a straightforward discretization:
e.g. local power series expansions do not make sense on a lattice,
so we cannot really speak of discrete analyticity.
On the other hand, as soon as we define discrete derivatives,
we can ask for the holomorphicity condition.
Thus it is philosophically more correct to speak of
\emph{discrete holomorphic},
rather than discrete analytic functions.
We will use the term \emph{preholomorphic} 
introduced by Ferrand \cite{ferrand-prehol},
as we prefer it to the term \emph{monodiffric}
used by Isaacs in the original papers \cite{isaacs,isaacs-52}
(a play on the term monogenic used by Cauchy for continuous analytic functions).

Though the preholomorphic functions are easy to define, there
is a lack of expository literature about them.
We see two main reasons:
firstly, there is no canonical preholomorphicity definition, and 
one can argue which of the competing approaches is better
(the answer probably depends on potential applications).
Secondly, it is straightforward to transfer to the discrete case
beginnings of the usual complex analysis
(a nice topic for an undergraduate research project),
but the easy life ends when it becomes necessary to multiply preholomorphic functions.
There is no easy and  natural way to proceed and the 
difficulty is addressed depending on the problem at hand.

As there seems to be no canonical discretization
of the complex analysis, we would rather adopt a utilitarian
approach, working with definitions
corresponding to interesting objects of probabilistic origin,
and allowing for a passage to the scaling limit.
We want to emphasize, that we are concerned with the following triplets:
\begin{enumerate}
  \item A planar graph,
  \item Its embedding into the complex plane, 
  \item Discrete Cauchy-Riemann equations.
\end{enumerate}  
We are interested in triplets such that the discrete complex analysis approximates
the continuous one.
Note that one can start with only a few elements of the triplet,
which gives some freedom.
For example, given an embedded graph, one can ask which discrete
difference equations have solutions close to holomorphic functions.
Or, given a planar graph and a notion of preholomorphicity, one can look for an 
appropriate embedding.

The ultimate goal is to find lattice models
of statistical physics
with preholomorphic observables. Since those observables
would approximate holomorphic functions,
some information about the original model could be 
subsequently deduced.

Below we start with several possible definitions of the
preholomorphic functions along with historical remarks.
Then we discuss some of their recent applications in probability
and statistical physics.

\section{Discrete holomorphic functions}\label{sec:history}

For a given planar graph, there are several ways to define
\emph{preholomorphic} functions,
and it is not always clear which way is preferable.
A much better known class is that of 
\emph{discrete harmonic} (or \emph{preharmonic}) functions,
which can be defined on any graph (not necessarily planar),
and also in more than one way. 
However, one definition stands out as the simplest:
a function on the vertices of graph is said to be \emph{preharmonic}
at a vertex $v$, if its discrete Laplacian vanishes:
\begin{equation}\label{eq:lapl}
0~=~\Delta H (u)~:=~\sum_{v:~\mathrm{neighbor~of~}u}\br{H(v)-H(u)}~.
\end{equation}
More generally, one can put weights on the edges, 
which would amount to taking different resistances in the
electric interpretation below.
Preharmonic functions on planar graphs are closely
related to discrete holomorphicity: for example,
their gradients defined on the oriented edges by
\begin{equation}\label{eq:grad}
F(\vec{uv})~:=~H(v)-H(u)~, 
\end{equation}
are preholomorphic.
Note that the edge function above is antisymmetric, i.e.
$F(\vec{uv})=-F(\vec{vu})$.

Both classes with the definitions as above 
are implicit already in the 1847 work of Kirchhoff \cite{kirchhoff-1847},
who interpreted a function defined on oriented edges
as an electric current flowing through the graph.
If we assume that all edges have unit resistance, than
the sum of currents flowing from a vertex is zero
by the first Kirchhoff law:
\begin{equation}\label{eq:kir1}
\sum_{u:~\mathrm{neighbor~of~}v}F\br{\vec{uv}}~=~0~,
\end{equation}
and the sum of the currents around any oriented closed contour $\gamma$
(for the planar graphs it is sufficient to consider contours around faces)
face is zero by the second Kirchhoff law:
\begin{equation}\label{eq:kir2}
\sum_{\vec{uv}\in\gamma}F\br{\vec{uv}}~=~0~.
\end{equation}
The two laws are equivalent to saying that
$F$ is given by the gradient of a potential function $H$ as in \eqref{eq:grad},
and the latter function is preharmonic \eqref{eq:lapl}.
One can equivalently think of a hydrodynamic interpretation,
with $F$ representing the flow of liquid.
Then conditions \eqref{eq:kir1} and \eqref{eq:kir2} mean that the flow
is divergence- and curl-free correspondingly.
Note that in the continuous setting  similarly defined 
gradients of harmonic functions on planar domains coincide up to complex conjugation with holomorphic
functions. And in higher dimensions harmonic gradients were proposed
 as one of their possible generalizations.


There are many other ways to introduce
discrete structures on graphs, which can be developed in
parallel to the usual complex analysis.
We have in mind mostly such discretizations 
that restrictions of holomorphic (or harmonic) functions become
\emph{approximately} preholomorphic (or preharmonic). Thus we speak about 
graphs embedded into the complex plane or a Riemann surface,
and the choice of embedding plays an important role.
Moreover, the applications we are after require passages to the scaling limit
(as mesh of the lattice tends to zero),
so we want to deal with discrete structures which converge to the usual 
complex analysis as we take finer and finer graphs.

Preharmonic functions satisfying \eqref{eq:lapl} on the square lattices
with decreasing mesh fit well into this philosophy,
and were studied in
a number of papers in early twentieth century
(see e.g. \cite{phillips-wiener,bouligand,lusternik1926}),
culminating in the seminal work
of Courant, Friedrichs and Lewy.
It was shown in \cite{courant-friedrichs-lewy} that
solution to the Dirichlet problem for a discretization of an elliptic operator
converges to the solution of the analogous continuous problem
as the mesh of the lattice tends to zero.
In particular, 
a preharmonic function
with given boundary values converges in the scaling limit
to a harmonic function
with the same boundary values in a rather strong
sense, including convergence of all partial derivatives.

Preholomorphic functions distinctively appeared for the first time in the papers
\cite{isaacs,isaacs-52} of Isaacs, where he proposed two definitions
(and called such functions ``monodiffric'').
A few papers of his and others followed, studying the first definition \eqref{eq:cr1},
which is asymmetric on the square lattice.
More recently the first definition was studied by Dynnikov and Novikov 
\cite{dynnikov-novikov-mmj} in the triangular lattice context,
where it becomes symmetric
(the triangular lattice is obtained from the square lattice by adding all the diagonals
in one direction). 

The second, symmetric, definition
was reintroduced by Ferrand,
who also discussed the passage to the scaling limit \cite{ferrand-prehol,ferrand-book}.
This was followed by extensive studies of Duffin and others, starting with
\cite{duffin}.

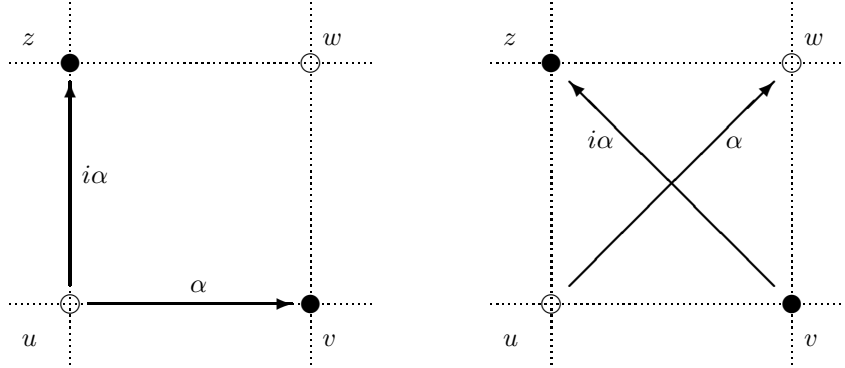
\begin{figure}
\unitlength=0.8mm
\begin{picture}(60,80)(-40,-10)
\put(50,0){{\thicklines
\put(13,13){\vector(1,1){34}}\put(47,13){\vector(-1,1){34}}}
\put(0,10){\qbezier[60](0,0)(30,0)(60,0)}
\put(0,50){\qbezier[60](0,0)(30,0)(60,0)}
\put(10,0){\qbezier[60](0,0)(0,30)(0,60)}
\put(50,0){\qbezier[60](0,0)(0,30)(0,60)}
\put(10,10){\circle{3}}\put(10,50){\circle*{3}}\put(50,10){\circle*{3}}\put(50,50){\circle{3}}
\put(39,36){${\alpha}$}\put(16,36){${i\alpha}$}
\put(2,3){${u}$}\put(52,3){${v}$}\put(52,53){${w}$}\put(2,53){${z}$}}
\put(-30,0){
\put(30,12){${\alpha}$}\put(12,30){${i\alpha}$}
{\thicklines
\put(13,10){\vector(1,0){34}}\put(10,13){\vector(0,1){34}}}
\put(0,10){\qbezier[60](0,0)(30,0)(60,0)}
\put(0,50){\qbezier[60](0,0)(30,0)(60,0)}
\put(10,0){\qbezier[60](0,0)(0,30)(0,60)}
\put(50,0){\qbezier[60](0,0)(0,30)(0,60)}
\put(10,10){\circle{3}}\put(10,50){\circle*{3}}\put(50,10){\circle*{3}}\put(50,50){\circle{3}}
\put(2,3){${u}$}\put(52,3){${v}$}\put(52,53){${w}$}\put(2,53){${z}$}}
\end{picture}
\caption{\label{fig:cr}
The first and the second Isaacs' definitions of discrete holomorphic
functions: multiplied by $i$ difference along the vector $\alpha$ is equal
to the difference along the rotated vector $i\alpha$.
Note that the second definition (on the right) is symmetric
with respect to lattice rotations, while the first one is not.
}
\end{figure}

Both definitions 
ask for a discrete version of the
Cauchy-Riemann equations ${\partial_{i\alpha}F}=i{\partial_{\alpha}F}$
or equivalently that $z$-derivative is independent of direction.
Consider a subregion $\Omega_\epsilon$ of the mesh $\epsilon$ square lattice
$\epsilon\ZZ^2\subset\CC$ and define a function on its vertices.
Isaacs proposed the following two definitions, 
replacing the derivatives by discrete differences.
His ``monodiffric functions of the first kind''
are required to satisfy inside 
$\Omega_\epsilon$ the following identity:
\begin{equation}\label{eq:cr1}
{F(z+i\epsilon)-F(z)}~=~i\br{{F(z+\epsilon)-F(z)}}~,
\end{equation}
which can be rewritten as
$$\frac{F(z+i\epsilon)-F(z)}{(z+i\epsilon)-z}~=
~\frac{{F(z+\epsilon)-F(z)}}{(z+\epsilon)-z}~.$$
We will be working with his second definition, 
which is more symmetric and also appears naturally in probabilistic
context (but otherwise the theories based on two definitions
are almost the same).
We say that a function
is \emph{preholomorphic}, if inside $\Omega_\epsilon$ it satisfies
the following identity, illustrated in Figure~\ref{fig:cr}:
\begin{equation}\label{eq:cr}
{F(z+i\epsilon)-F(z+\epsilon)}~=~i\br{{F(z+\epsilon(1+i))-F(z)}}~,
\end{equation}
which can also be rewritten as
$$\frac{F(z+i\epsilon)-F(z+\epsilon)}{(z+i\epsilon)-(z+\epsilon)}~=
~\frac{{F(z+\epsilon(1+i))-F(z)}}{(z+\epsilon(1+i))-z}~.$$
It is easy to see that restrictions of continuous
holomorphic functions to the mesh $\epsilon$ square lattice
satisfy this identity up to $O(\epsilon^3)$.
Note also that if we color the lattice in
the chess-board fashion,
the complex identity \eqref{eq:cr}
can be written as two real identities (its real and imaginary parts),
one involving the real part
of $F$ at  black vertices and the imaginary part
of $F$ at white vertices, the other one -- vice versa.
So unless we have special boundary conditions,
$F$ splits into two ``demi-functions''
(real at white and imaginary at black vs. 
imaginary at black and real at white vertices),
and some prefer to consider just one of those,
i.e. ask $F$ to be purely real at black vertices
and purely imaginary at white ones.

The theory of so defined preholomorphic functions starts much
like the usual complex analysis.
It is easy to check, that for preholomorphic functions
sums are also preholomorphic,
discrete contour integrals vanish,
primitive (in a simply-connected domain) and derivative
are well-defined and are preholomorphic functions on the dual square lattice,
real and imaginary parts are preharmonic 
on their respective black and white sublattices,
etc.
Unfortunately, the product of two preholomorphic functions
is no longer preholomorphic:
e.g., while restrictions of $1$, $z$, and $z^2$
to the square lattice are preholomorphic, the higher powers are only approximately so.

Situation with other possible definitions is similar,
with much of the linear complex analysis being easy to reproduce,
and problems appearing when one has to multiply preholomorphic functions.
Pointwise multiplication cannot be consistently defined, and though
one can introduce convolution-type multiplication,
the possible constructions are non-local and cumbersome.
Sometimes, for different graphs and definitions, problems appear even earlier, 
with the first derivative not being preholomorphic.

Our main reason for choosing the definition \eqref{eq:cr}
is that it naturally appears in probabilistic context.
It was also noticed by Duffin that \eqref{eq:cr} 
nicely generalizes to a larger family of 
\emph{rhombic lattices}, where all the faces are rhombi.
Equivalently, one can speak of \emph{isoradial graphs},
where all faces are inscribed into circles of the same radius
--- an isoradial graph together with its dual
forms a rhombic lattice.

There are two main reasons to study this  particular family.
First, this is perhaps the largest family of graphs for which the Cauchy-Riemann operator admits a nice discretization.
Indeed, restrictions of holomorphic functions to such
graphs are preholomorphic to higher orders.
This was the reason for the introduction of complex analysis on rhombic lattices 
by Duffin \cite{duffin-rhombic} in late sixties.
More recently, the complex analysis on such graphs was studied for
the sake of probabilistic applications
\cite{mercat-2001,kenyon-operators,chelkak-smirnov-dca}.

On the other hand, this seems to be the largest family where certain lattice models,
including the Ising model, have nice integrability properties.
In particular, the critical point can be defined
with weights depending only on the local structure,
and the star-triangle relation works out nicely.
It seems that the first appearance of related family of graphs in
the probabilistic context
was in the work of Baxter \cite{baxter-ptrsl-1978},
where the eight vertex and Ising models
were considered on $Z$-invariant graphs, arising from planar line arrangements.
These graphs are topologically the same as the isoradial ones,
and though they are embedded differently into the plane,
by \cite{kenyon-schlenker}  they always admit isoradial embeddings.
In \cite{baxter-ptrsl-1978} Baxter was not passing to the scaling limit,
and so the actual choice of embedding was immaterial 
for his results.
However, his choice of weights in the models would 
suggest an isoradial embedding, and the Ising model 
was so considered
by Mercat \cite{mercat-2001},
Boutilier and de~Tili\`ere
\cite{boutillier-tiliere-periodic,boutillier-tiliere-locality},
Chelkak and the author 
\cite{chelkak-smirnov-iso}.
Additionally, the dimer and the uniform spanning tree models on such
graphs also have nice properties, see e.g. \cite{kenyon-operators}.

We would also like to remark that rhombic lattices form a rather large family of graphs.
While not every topological quadrangulation
(graph all of whose faces are quadrangles)
admits a rhombic embedding,
Kenyon and Schlenker \cite{kenyon-schlenker} 
gave a simple topological condition 
necessary and sufficient for its existence.

So this seems to be the most general family of graphs appropriate
for our subject, and most of what we discuss below generalizes to it
(though for simplicity we speak of the square and hexagonal lattices only).

\section{Applications of preholomorphic functions}\label{sec:applications}

Besides being interesting in themselves,
preholomorphic functions found several diverse applications in 
combinatorics, analysis,
geometry, probability and physics.

After the original work
of Kirchhoff, the first notable application was perhaps the 
famous article  \cite{brooks-squares} 
of Brooks, Smith, Stone and Tutte, who used 
preholomorphic functions to construct tilings of
rectangles by squares.

Several applications to analysis followed,
starting with a new proof of the Riemann uniformization theorem by
Ferrand \cite{ferrand-book}.
Solving the discrete version of the usual minimization problem, it is 
immediate to establish the existence of the minimizer and its properties,
and  then one shows that it has a scaling limit,
which is the desired uniformization.
Duffin and his co-authors found a number of
similar applications,
including construction of the Bergman kernel by
Dieter and Mastin \cite{deeter-mastin}. 
There were also studies of discrete
versions of the multi-dimensional complex analysis,
see e.g. Kiselman's \cite{kiselman}.

In \cite{thurston-zippers} Thurston 
proposed \emph{circle packings} as another discretization
of complex analysis.
They found some beautiful applications,
including yet another proof
of the Riemann uniformization theorem
by Rodin and Sullivan \cite{rodin-sullivan}.
More interestingly, they were used by He and Schramm \cite{he-schramm-fixed}
in the best result so far on the Koebe uniformization conjecture,
stating that any domain can be conformally uniformized
to a domain bounded by circles and points.
In particular, they established the conjecture for domains with
countably many boundary components.
More about circle packings
can be learned form Stephenson's book \cite{stephenson-book}.
Note that unlike the discretizations discussed above, 
the circle packings lead to non-linear versions
of the Cauchy-Riemann equations, 
see e.g. the discussion in \cite{bobenko-mercat-suris}.

There are other interesting applications to geometry, analysis, combinatorics,
probability, and we refer the interested reader to the expositions
by Lov\'asz \cite{lovasz-dca}, Stephenson \cite{stephenson-book},
Mercat \cite{mercat-drs}, Bobenko and Suris 
\cite{bobenko-suris-book}.


In this note we are interested in applications to probability and
statistical physics.
Already the Kirchhoff's paper \cite{kirchhoff-1847} makes connection
between the Uniform Spanning Tree and  preharmonic (and so preholomorphic)
functions.

Connection of Random Walk to preharmonic functions
was certainly known to many researchers in early
twentieth century, and
figured implicitly in many papers.
It is explicitly discussed by Courant, Friedrichs and Lewy
in \cite{courant-friedrichs-lewy},
with preharmonic functions appearing as Green's functions 
and exit probabilities for the Random Walk.

More recently, Kenyon found preholomorphic
functions in the dimer model
(and in the Uniform Spanning Tree in a way different
from the original considerations of Kirchhoff).
He was able to obtain many beautiful results about statistics
of the dimer tilings, and in particular,
showed that those have a conformally invariant
scaling limit, described by the Gaussian Free Field,
see \cite{kenyon-conformal,kenyon-gff}.
More about Kenyon's results can be found in his expositions
\cite{kenyon-introduction,kenyon-lectures}.
An approximately preholomorphic function was found
by the author in the critical site percolation on
the triangular lattice, allowing to prove the Cardy's
formula for crossing probabilities \cite{smirnov-cras,smirnov-perc}.

Finally, we remark that various other discrete relations were observed in many
integrable two dimensional models of statistical physics,
but usually no explicit connection was made with complex analysis,
and no scaling limit was considered.
Here we are interested in applications of integrability
parallel to that for the Random Walk and the dimer model above.
Namely, once a preholomorphic function
is observed in some probabilistic model,
we can pass to the scaling limit, obtaining
a holomorphic function.
Thus, the preholomorphic observable 
is approximately equal to the limiting holomorphic function,
providing some knowledge about the model at hand.
Below we discuss applications of this philosophy,
starting with the Ising model.


\section{The Ising model}\label{sec:ising}

In this Section we discuss some of the ways
how preholomorphic functions appear in the Ising model at criticality.
The observable below was proposed in \cite{smirnov-icm}
for the hexagonal lattice, along with a possible
generalization to $O(N)$ model.
Similar objects appeared earlier in Kadanoff and Ceva \cite{kadanoff-ceva} 
and in Mercat \cite{mercat-2001}, though
boundary values and conformal covariance, which are central to us, were never discussed.

The scaling limit and properties of our observable
on isoradial graphs were worked out
by Chelkak and the author in 
\cite{chelkak-smirnov-iso}.
It is more appropriate to consider it as a \emph{fermion}
or a \emph{spinor}, by writing
$F(z)\sqrt{dz}$, 
and with more general setup one has to proceed in this way.

Earlier we constructed a similar fermion
for the random cluster representation of the Ising model,
see \cite{smirnov-icm,smirnov-fk1} and 
our joint work with Chelkak 
\cite{chelkak-smirnov-iso} for generalization to isoradial graphs 
(and also independent work of Riva and Cardy 
\cite{riva-cardy-hol} for its physical connections).
It has a simpler probabilistic interpretation than the
fermion in the spin representation,
as it can be written as the probability
of the interface  between two marked boundary points 
passing through a point inside, corrected by a complex weight
depending on the winding.

The fermion for the spin representation is more difficult to construct.
Below we describe it in terms of contour collections with distinguished
points.
Alternatively it corresponds to the partition function of the Ising model with
a $\sqrt{z}$ monodromy at a given edge, corrected by a complex weight;
or to a product of order and disorder operators
at neighboring site and dual site.

We will consider the Ising model on the mesh $\epsilon$ square lattice.
Let $\Omega_\epsilon$ be a discretization of some
bounded domain $\Omega\subset\CC$.
The Ising model on $\Omega_\epsilon$
has configurations $\sigma$ which assign $\pm1$ (or simply $\pm$) spins $\sigma(v)$ to 
 vertices $v\in\Omega_\epsilon$ and
Hamiltonian defined (in the absence of an external magnetic field) by
$$H(\sigma)~=~-\sum_{\brl{u,v}}~\sigma(u)\sigma(v)~,$$
where the sum is taken over all edges $\brl{u,v}$ inside $\Omega_\epsilon$.
Then the partition function is given by
$$Z~=~\sum_{\sigma}~\exp\br{-\beta H(\sigma)}~,$$
and probability of a given spin configuration becomes
$$\PP\br{\sigma} ~=~\exp\br{-\beta H(\sigma)}/Z~.$$

Here $\beta\ge0$ is the temperature parameter
(behaving like the reciprocal of the actual temperature),
and Kramers and Wannier have established \cite{kramers-wannier-i} that its critical value is 
given by $\beta_c=\log\br{\sqrt2+1}/2$.

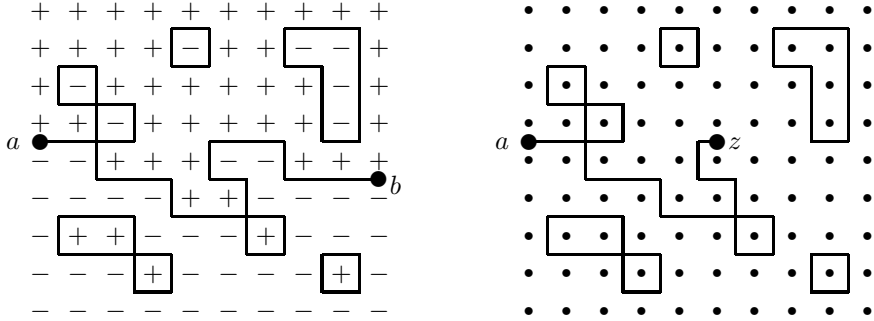
\begin{figure}
\unitlength=1.mm
\thicklines
\begin{picture}(120,40)
\put(75,5){\def\vtx{\circle*{2}}
\def\rr{\line(1,0){5}}\def\uu{\line(0,1){5}}
\def\pp{\circle*{1}}\def\mm{\circle*{1}}
\put(2.5,2.5){
\put(-5,-5)\mm\put(0,-5)\mm\put(5,-5)\mm\put(10,-5)\mm\put(15,-5)\mm\put(20,-5)\mm\put(25,-5)\mm\put(30,-5)\mm\put(35,-5)\mm\put(40,-5)\mm
\put(-5,0)\mm\put(0,0)\mm\put(5,0)\mm\put(10,0)\pp\put(15,0)\mm\put(20,0)\mm\put(25,0)\mm\put(30,0)\mm\put(35,0)\pp\put(40,0)\mm
\put(-5,5)\mm\put(0,5)\pp\put(5,5)\pp\put(10,5)\mm\put(15,5)\mm\put(20,5)\mm\put(25,5)\pp\put(30,5)\mm\put(35,5)\mm\put(40,5)\mm
\put(-5,10)\mm\put(0,10)\mm\put(5,10)\mm\put(10,10)\mm\put(15,10)\pp\put(20,10)\pp\put(25,10)\mm\put(30,10)\mm\put(35,10)\mm\put(40,10)\mm
\put(-5,15)\mm\put(0,15)\mm\put(5,15)\pp\put(10,15)\pp\put(15,15)\pp\put(20,15)\mm\put(25,15)\mm\put(30,15)\pp\put(35,15)\pp\put(40,15)\pp
\put(-5,20)\pp\put(0,20)\pp\put(5,20)\mm\put(10,20)\pp\put(15,20)\pp\put(20,20)\pp\put(25,20)\pp\put(30,20)\pp\put(35,20)\mm\put(40,20)\pp
\put(-5,25)\pp\put(0,25)\mm\put(5,25)\pp\put(10,25)\pp\put(15,25)\pp\put(20,25)\pp\put(25,25)\pp\put(30,25)\pp\put(35,25)\mm\put(40,25)\pp
\put(-5,30)\pp\put(0,30)\pp\put(5,30)\pp\put(10,30)\pp\put(15,30)\mm\put(20,30)\pp\put(25,30)\pp\put(30,30)\mm\put(35,30)\mm\put(40,30)\pp
\put(-5,35)\pp\put(0,35)\pp\put(5,35)\pp\put(10,35)\pp\put(15,35)\pp\put(20,35)\pp\put(25,35)\pp\put(30,35)\pp\put(35,35)\pp\put(40,35)\pp
}
\put(10,0)\rr\put(10,5)\rr\put(5,5)\rr\put(5,10)\rr
\put(0,5)\rr\put(0,10)\rr\put(15,30)\rr\put(15,35)\rr
\put(30,30)\rr\put(35,35)\rr\put(35,0)\rr\put(35,5)\rr\put(35,20)\rr
\put(30,35)\rr
{\red\put(5,25)\uu\put(30,5)\uu
\put(25,10)\rr\put(25,5)\rr\put(-2.5,20){\line(1,0){7.5}}\put(0,25)\rr\put(0,30)\rr
\put(5,15)\rr\put(5,20)\rr\put(5,25)\rr
\put(10,15)\rr\put(15,10)\rr\put(20,10)\rr\put(20,15)\rr\put(20,20){\line(1,0){2.5}}
\put(0,25)\uu\put(5,15)\uu\put(5,20)\uu
\put(10,20)\uu\put(15,10)\uu\put(20,15)\uu\put(25,5)\uu\put(25,10)\uu}
\put(15,0)\uu\put(0,5)\uu\put(10,0)\uu\put(10,5)\uu\put(15,30)\uu
\put(30,30)\uu\put(35,0)\uu\put(35,20)\uu\put(35,25)\uu
\put(40,0)\uu\put(40,20)\uu\put(40,25)\uu\put(40,30)\uu\put(20,30)\uu
\put(-2.5,20)\vtx\put(22.5,20)\vtx\put(-7,19){$a$}\put(24,19){$z$}}
\put(10,5)
{\def\vtx{\circle*{2}}\def\rr{\line(1,0){5}}\def\uu{\line(0,1){5}}
\def\pp{$+$}\def\mm{$-$}
\put(1.2,1.5){
\put(-5,-5)\mm\put(0,-5)\mm\put(5,-5)\mm\put(10,-5)\mm\put(15,-5)\mm\put(20,-5)\mm\put(25,-5)\mm\put(30,-5)\mm\put(35,-5)\mm\put(40,-5)\mm
\put(-5,0)\mm\put(0,0)\mm\put(5,0)\mm\put(10,0)\pp\put(15,0)\mm\put(20,0)\mm\put(25,0)\mm\put(30,0)\mm\put(35,0)\pp\put(40,0)\mm
\put(-5,5)\mm\put(0,5)\pp\put(5,5)\pp\put(10,5)\mm\put(15,5)\mm\put(20,5)\mm\put(25,5)\pp\put(30,5)\mm\put(35,5)\mm\put(40,5)\mm
\put(-5,10)\mm\put(0,10)\mm\put(5,10)\mm\put(10,10)\mm\put(15,10)\pp\put(20,10)\pp\put(25,10)\mm\put(30,10)\mm\put(35,10)\mm\put(40,10)\mm
\put(-5,15)\mm\put(0,15)\mm\put(5,15)\pp\put(10,15)\pp\put(15,15)\pp\put(20,15)\mm\put(25,15)\mm\put(30,15)\pp\put(35,15)\pp\put(40,15)\pp
\put(-5,20)\pp\put(0,20)\pp\put(5,20)\mm\put(10,20)\pp\put(15,20)\pp\put(20,20)\pp\put(25,20)\pp\put(30,20)\pp\put(35,20)\mm\put(40,20)\pp
\put(-5,25)\pp\put(0,25)\mm\put(5,25)\pp\put(10,25)\pp\put(15,25)\pp\put(20,25)\pp\put(25,25)\pp\put(30,25)\pp\put(35,25)\mm\put(40,25)\pp
\put(-5,30)\pp\put(0,30)\pp\put(5,30)\pp\put(10,30)\pp\put(15,30)\mm\put(20,30)\pp\put(25,30)\pp\put(30,30)\mm\put(35,30)\mm\put(40,30)\pp
\put(-5,35)\pp\put(0,35)\pp\put(5,35)\pp\put(10,35)\pp\put(15,35)\pp\put(20,35)\pp\put(25,35)\pp\put(30,35)\pp\put(35,35)\pp\put(40,35)\pp
}
\put(10,0)\rr\put(10,5)\rr\put(5,5)\rr\put(5,10)\rr
\put(0,5)\rr\put(0,10)\rr\put(15,30)\rr\put(15,35)\rr
\put(30,30)\rr\put(35,35)\rr\put(35,0)\rr\put(35,5)\rr\put(35,20)\rr
\put(30,35)\rr
{\red\put(5,25)\uu\put(30,5)\uu
\put(25,10)\rr\put(25,5)\rr\put(-2.5,20){\line(1,0){7.5}}\put(0,25)\rr\put(0,30)\rr
\put(5,15)\rr\put(5,20)\rr\put(5,25)\rr
\put(10,15)\rr\put(15,10)\rr\put(20,10)\rr\put(20,15)\rr
\put(20,20)\rr\put(25,20)\rr
\put(30,15)\uu\put(30,15){\line(1,0){12.5}}
\put(0,25)\uu\put(5,15)\uu\put(5,20)\uu
\put(10,20)\uu\put(15,10)\uu\put(20,15)\uu\put(25,5)\uu\put(25,10)\uu}
\put(15,0)\uu\put(0,5)\uu\put(10,0)\uu\put(10,5)\uu\put(15,30)\uu
\put(30,30)\uu\put(35,0)\uu\put(35,20)\uu\put(35,25)\uu
\put(40,0)\uu\put(40,20)\uu\put(40,25)\uu\put(40,30)\uu\put(20,30)\uu
\put(-2.5,20)\vtx\put(42.5,15)\vtx\put(-7,19){$a$}
\put(44,13){$b$}}
\end{picture}
\caption{Left: configuration of spins in the Ising model
with Dobrushin boundary conditions, its contour 
representation, and an interface between two boundary points.
Right: an example of a configuration considered for the Fermionic observable:
a number of loops and a contour connecting $a$ to $z$. It can be represented as a spin
configuration with a monodromy at $z$.
\label{fig:ising}}
\end{figure}

Now represent the spin configurations  graphically by a collection of interfaces --
contours on the dual lattice, separating plus spins from minus spins,
the so-called \emph{low-temperature expansion},
see Figure~\ref{fig:ising}.
A contour collection is a set of edges, such that an even number emanates from every vertex.
In such case the contours can be represented as a union of loops
(possibly in a non-unique way, but we do not distinguish between different representations).
Note that each contour collection 
corresponds to two spin collections which
are negatives of each other, or to one if we fix the spin value at some vertex.
The partition function of the Ising model can be rewritten in terms
of the contour configurations $\omega$ as
$$Z~=~\sum_{\omega}x^{\mathrm{length~of~contours}}~.$$
Each neighboring pair of opposite spins contributes an edge to the contours,
and so a factor of $x=\exp(-2\beta)$ to the partition function.
Note that the critical value is $x_c=\exp(-2\beta_c)=\sqrt2-1$.

We now want to define a preholomorphic observable.
To this effect we need to distinguish at least one point
(so that the domain has a non-trivial conformal modulus).
One of the possible applications lies in relating interfaces to 
Schramm's SLE curves, in the simplest setup running between
two boundary points. To obtain a discrete interface between
two boundary points $a$ and $b$, we introduce
Dobrushin boundary conditions: $+$ on one boundary arc and $-$ on another,
see Figure~\ref{fig:ising}. Then those become unique points with 
an odd number of contour edges emanating from them.

Now to define our fermion, we allow the second endpoint of the interface
to move inside the domain.
Namely, take an edge center $z$ inside $\Omega_\epsilon$,
and define
\begin{equation}\label{eq:fising}
F_\epsilon(z)~:=~\sum_{\omega(a\to z)} x^{\mathrm{length~of~contours}}\,{\cal W}(\omega(a\to z))~,
\end{equation}
where the sum is taken over all contour configurations 
$\omega=\omega(a\to z)$ which have two exceptional points:
$a$ on the boundary and $z$ inside.
So the contour collection can be represented 
(perhaps non-uniquely)
as a collection of loops
plus an interface between $a$ and $z$.

\begin{figure}
\def\vtx{\circle*{3}}
\unitlength=0.7mm
\begin{picture}(180,90)(-10,0)
\thicklines
\put(40,70){
\black\put(-15,7){\oval(10,5)}\black\put(14,8){\oval(7,7)}\black\put(20,-5){\oval(5,12)}
\red\qbezier(-30,0)(0,0)(0,0)
\black\put(0,0){\oval(60,30)}\put(-30,0)\vtx\put(0,0)\vtx
\put(-37,-7){$a$}\put(-3,-7){$z$}
\put(-30,20){winding$=0$}\put(10,20){${\cal W}=1$}}
\put(120,70){
\black\put(-15,7){\oval(10,5)}\black\put(14,8){\oval(7,7)}\black\put(20,-5){\oval(5,12)}
\red\qbezier(-18,0)(-30,0)(-30,0)
\qbezier(-10,-5)(-10,0)(-18,0)
\qbezier(-10,-5)(-10,-10)(0,-10)
\qbezier(0,-10)(5,-10)(5,-5)
\qbezier(5,-5)(5,0)(0,0)
\black\put(0,0){\oval(60,30)}\put(-30,0)\vtx\put(0,0)\vtx
\put(-37,-7){$a$}\put(-3,-7){$z$}
\put(-30,20){winding$=\pi$}\put(10,20){${\cal W}=-i$}}
\put(40,20){
\black\put(-15,7){\oval(10,5)}\black\put(14,8){\oval(7,7)}\black\put(20,-5){\oval(5,12)}
\red\qbezier(-30,0)(-30,0)(-6,0)
\qbezier(-6,0)(-5,0)(-5,-1)
\qbezier(-5,5)(-5,1)(-5,1)
\qbezier(-5,-1)(-5,-13)(0,-13)
\qbezier(0,-13)(10,-13)(10,0)
\qbezier(10,0)(10,10)(0,10)
\qbezier(0,10)(-5,10)(-5,5)
\qbezier(-5,1)(-5,0)(-4,0)
\qbezier(-4,0)(-4,0)(0,0)
\black\put(-30,0)\vtx\put(0,0){\oval(60,30)}
\put(0,0)\vtx\put(-37,-7){$a$}\put(-1,-7){$z$}
\put(-30,20){winding$=\pm2\pi$}\put(10,20){${\cal W}=-1$}}
\put(120,20){
\black\put(-15,7){\oval(10,5)}\black\put(14,8){\oval(7,7)}\black\put(20,-5){\oval(5,12)}
\red\qbezier(-30,0)(-30,0)(-23,0)
\qbezier(-23,0)(-18,0)(-15,-7.5)
\qbezier(-15,-7.5)(-12,-13)(0,-13)
\qbezier(0,-13)(10,-13)(10,0)
\qbezier(10,0)(10,10)(0,10)
\qbezier(0,10)(-5,10)(-5,5)
\qbezier(-5,5)(-5,-3)(0,-3)
\qbezier(0,-3)(4,-3)(4,-2)
\qbezier(0,0)(4,0)(4,-2)
\black\put(-30,0)\vtx\put(0,0){\oval(60,30)}\put(0,0)\vtx
\put(-37,-7){$a$}\put(-3,-7){$z$}
\put(-30,20){winding$=3\pi$}\put(10,20){${\cal W}=i$}}
\end{picture}
\caption{\label{fig:winding}
Examples of Fermionic weights one obtains depending on the winding of the interface.
Note that in the bottom left example there are two ways to trace the interface from $a$
to $z$ without self-intersections, which give different windings
$\pm2\pi$, but the same complex weight ${\cal W}=-1$.}
\end{figure}
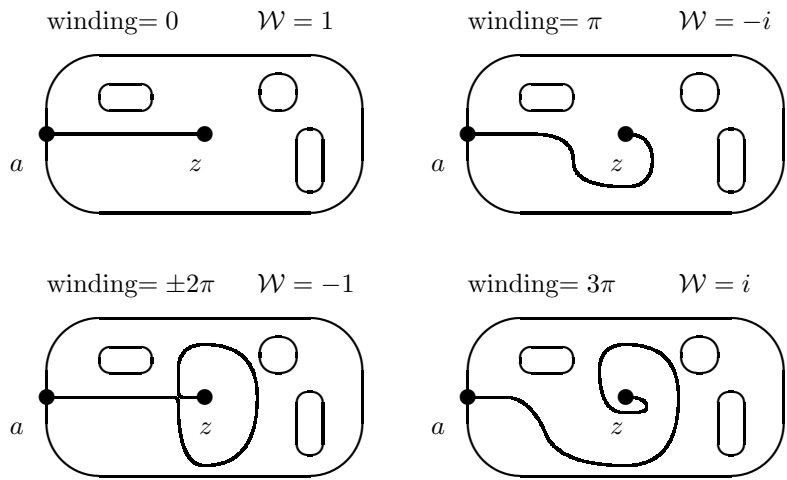

Furthermore, the sum is corrected 
by a \emph{Fermionic} complex weight, depending on the configuration:
$${\cal W}(\omega(a\to z))~:=~\exp\br{-{i}~\spin~\mathrm{winding}(\gamma, a\to z)}~.$$
Here the winding is the total turn of the interface $\gamma$ connecting $a$ to $z$,
counted in radians, and the \emph{spin} $\spin$ is equal to $1/2$
(it should not be confused with the Ising spins $\pm1$). 
For some collections the interface can be chosen in more than one way,
and then we trace it by taking a left turn whenever an ambiguity arises.
Another choice might lead to a different value of winding, but
if the loops and the interface have no ``transversal'' self-intersections, 
then the difference will be a multiple of $4\pi$
and so the complex weight ${\cal W}$ is well-defined.
Equivalently we can write
$${\cal W}(\omega(a\to z))~=
~\lambda^{\# \mathrm{~signed~turns~of~}\gamma}~,
~~~\lambda~:=~\exp\br{-i\spin\frac{\pi}{2}}~,$$
see Figure~\ref{fig:winding} for weights corresponding to different windings.

\begin{remark}
Removing complex weight $\cal W$ one retrieves the correlation of spins 
on the dual lattice at the dual temperature $x^*$,
a corollary of the Kramers-Wannier duality.
\end{remark}

\begin{remark}
While such contour collections cannot be directly 
represented by spin configurations,
one can obtain them
by creating a \emph{disorder operator}, i.e. a
\emph{monodromy} at $z$: when one goes one time around $z$,
spins change their signs.
\end{remark}

Our first theorem is the following, 
which is proved for general isoradial graphs in \cite{chelkak-smirnov-iso},
with a shorter proof for the square lattice given in \cite{chelkak-smirnov-spin}:
\begin{theorem}[Chelkak, Smirnov]\label{thm:spin}
For Ising model at criticality,
$F$ is a preholomorphic solution of
a Riemann boundary value problem.
When mesh $\epsilon\to0$, 
$${F_\epsilon(z)}\,/\,\sqrt{{\epsilon}}~\unif~ \sqrt{P'(z)}~~\mathit{{inside}}~\Omega,$$
where $P$ is the complex Poisson kernel at $a$:
a conformal map $\Omega\to\CC_+$ such that $a\mapsto\infty$.
Here both sides should be normalized in the same chart around $b$.
\end{theorem}

\begin{remark} 
For non-critical values of $x$ observable $F$ becomes \emph{massive preholomorphic},
satisfying the discrete analogue of the massive Cauchy-Riemann
equations: $\bar\partial\,F= im(x-x_c) \bar F$, cf. \cite{makarov-smirnov-icmp}.
\end{remark}

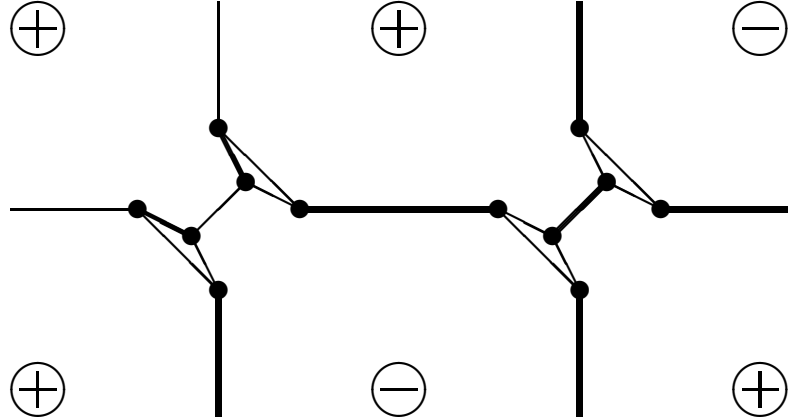
\begin{figure}
\unitlength=1.2mm
\begin{picture}(60,50)(-35,-30)
\thicklines
\put(-20,20){\put(0,0){\circle{6}}\put(-2,0){\line(1,0){4}}\put(0,-2){\line(0,1){4}}}
\put(-20,-20){\put(0,0){\circle{6}}\put(-2,0){\line(1,0){4}}\put(0,-2){\line(0,1){4}}}
\put(20,20){\put(0,0){\circle{6}}\put(-2,0){\line(1,0){4}}\put(0,-2){\line(0,1){4}}}
\put(20,-20){\put(0,0){\circle{6}}\put(-2,0){\line(1,0){4}}}
\put(60,20){\put(0,0){\circle{6}}\put(-2,0){\line(1,0){4}}}
\put(60,-20){\put(0,0){\circle{6}}\put(-2,0){\line(1,0){4}}\put(0,-2){\line(0,1){4}}}
\put(0,0){
\put(-9,0){\circle*{2}}\put(9,0){\circle*{2}}
\put(0,9){\circle*{2}}\put(0,-9){\circle*{2}}
\put(-3,-3){\circle*{2}}\put(3,3){\circle*{2}}
\put(-9,0){\line(1,-1){9}}\put(0,9){\line(1,-1){9}}
\put(-9.3,0){\line(2,-1){6}}\put(-8.7,0){\line(2,-1){6}}\put(-9,0){\line(2,-1){6}}
\put(0,-9){\line(-1,2){3}}\put(9,0){\line(-2,1){6}}
\put(0.2,9){\line(1,-2){3}}\put(-0.2,9){\line(1,-2){3}}\put(0,9){\line(1,-2){3}}
\put(-3,-3){\line(1,1){6}}
\put(9,0){\line(1,0){22}}\put(9,0.2){\line(1,0){22}}\put(9,-0.2){\line(1,0){22}}
\put(-9,0){\line(-1,0){14}}
\put(49,0){\line(1,0){14}}\put(49,-0.2){\line(1,0){14}}\put(49,0.2){\line(1,0){14}}
\put(0,9){\line(0,1){14}}
\put(0,-9){\line(0,-1){14}}\put(0.2,-9){\line(0,-1){14}}\put(-0.2,-9){\line(0,-1){14}}
\put(40.2,9){\line(0,1){14}}\put(40,9){\line(0,1){14}}\put(39.8,9){\line(0,1){14}}
\put(40,-9){\line(0,-1){14}}\put(40.2,-9){\line(0,-1){14}}\put(39.8,-9){\line(0,-1){14}}
}
\put(40,0){
\put(-9,0){\circle*{2}}\put(9,0){\circle*{2}}
\put(0,9){\circle*{2}}\put(0,-9){\circle*{2}}
\put(-3,-3){\circle*{2}}\put(3,3){\circle*{2}}
\put(-9,0){\line(1,-1){9}}\put(0,9){\line(1,-1){9}}
\put(-9,0){\line(2,-1){6}}\put(0,-9){\line(-1,2){3}}
\put(9,0){\line(-2,1){6}}\put(0,9){\line(1,-2){3}}
\put(-3,-3){\line(1,1){6}}\put(-2.7,-3){\line(1,1){6}}\put(-3.3,-3){\line(1,1){6}}
}
\end{picture}
\caption{\label{fig:fisher}
Fisher graph for a region of the square lattice, 
a spin configuration and a corresponding dimer configuration,
with dimers represented by the bold edges.}
\end{figure}

\begin{remark}
Ising model can be represented as a dimer model on the Fisher graph.
For example, on the square lattice, one first represents the spin configuration
as above ---
by the collection of contours
on the dual lattice, separating $+$ and $-$ spins.
Then the dual lattice is modified with every vertex replaced by
a ``city'' of six vertices, see Figure~\ref{fig:fisher}.
It is easy to see that there is a natural bijection
between contour configurations on the dual square lattice
and dimer configuration on its Fisher graph.

Then, similarly to the work of Kenyon for the square lattice,
the coupling function for the Fisher lattice
will satisfy difference equations,
which upon examination turn out to be another discretization
of Cauchy-Riemann equations,
with different projections of the preholomorphic
function assigned to six vertices in a ``city''.
One can then reinterpret the coupling function in terms of the
Ising model, and this is the approach taken by
Boutilier and de~Tili\`ere
\cite{boutillier-tiliere-periodic,boutillier-tiliere-locality}.

This is also how the author found the observable discussed in this Section,
observing jointly with Kenyon in 2002 that it has the potential
to imply the convergence of the interfaces to the Schramm's SLE curve.
\end{remark}

The key  to establishing Theorem~\ref{thm:spin} 
is the observation that the function $F$ is preholomorphic.
Moreover, it turns out that $F$ satisfies a stronger form of preholomorphicity,
which implies the usual one, but is better adapted to fermions.

Consider the function $F$ on the centers of edges.
We say that $F$ is \emph{strongly} (or \emph{spin}) preholomorphic
if for every  centers $u$ and $v$ of two neighboring edges emanating
from a vertex $w$,
we have
$$\mathrm{Proj}(F(v),1/\sqrt{\alpha})=\mathrm{Proj}(F(u),1/\sqrt{\alpha})~,$$ 
where $\alpha$ is the unit bisector of the angle $uwv$, and
$\mathrm{Proj}(p,q)$ denotes the orthogonal projection of
the vector $p$ on the vector $q$.
Equivalently we can write
\begin{equation}
F(v)+\bar\alpha\, \overline{F(v)}=F(u)+\bar\alpha\, \overline{F(u)}~.
\label{eq:shol}\end{equation}
This definition implies the classical one for the square lattice,
and it also easily adapts to the isoradial graphs.
Note that for convenience we assume that the interface starts from $a$
in the positive real direction as in Figure~\ref{fig:ising}, which slightly changes
weights compared to the convention in \cite{chelkak-smirnov-iso}.

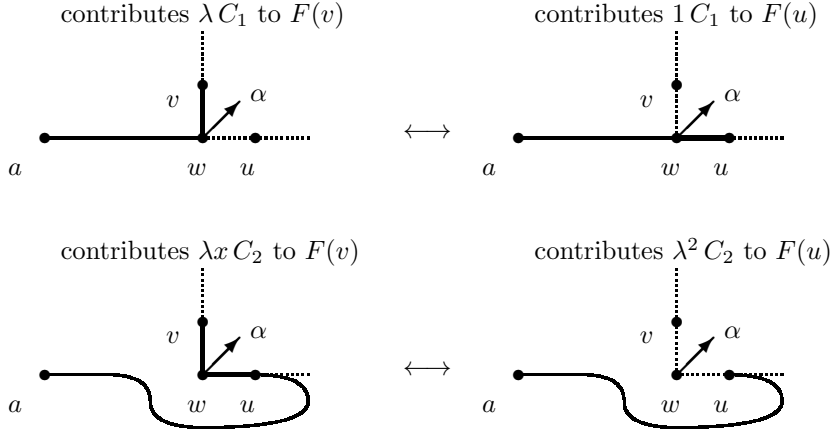
\begin{figure}
\def\vtx{\black\circle*{2}}
\def\vty{{\red\circle*{2}}}
\unitlength=0.7mm
\begin{picture}(220,100)(-10,-40)
\thicklines
\put(40,25){
\put(-27,22){contributes $\blue\lambda\,C_1$ to $\blue F(v)$}
\black
\red\qbezier(-30,0)(0,0)(0,0)
\qbezier(0,10)(0,0)(0,0)
\red\qbezier(-0.2,10)(-0.2,0)(-0.2,0)
\red\qbezier(0.2,10)(0.2,0)(0.2,0)
\black\qbezier[10](0,20)(0,15)(0,10)\qbezier[20](20,0)(10,0)(0,0)
\put(0,0){\blue\vector(1,1){7}}\put(9,7){\blue$\alpha$}
\black
\put(-30,0)\vtx\put(0,0)\vtx\put(10,0)\vty\put(0,10)\vty
\put(-37,-7){$a$}\put(-3,-7){$w$}\put(7,-7){$u$}\put(-7,6){$v$}
\put(38,0){$\red\longleftrightarrow$}
}
\put(130,25){
\black
\red\qbezier(-30,0)(0,0)(0,0)
\red\qbezier(10,0)(0,0)(0,0)
\qbezier(10,0.2)(0,0.2)(0,0.2)
\qbezier(10,-0.2)(0,-0.2)(0,-0.2)
\black\qbezier[20](0,20)(0,10)(0,0)\qbezier[10](20,0)(15,0)(10,0)
\put(0,0){\blue\vector(1,1){7}}\put(9,7){\blue$\alpha$}
\black
\put(-30,0)\vtx\put(0,0)\vtx\put(10,0)\vty\put(0,10)\vty
\put(-37,-7){$a$}\put(-3,-7){$w$}\put(7,-7){$u$}\put(-7,6){$v$}
\put(-27,22){contributes $\blue1\,C_1$ to $\blue F(u)$}
}
\put(40,-20){
\red\qbezier(-18,0)(-30,0)(-30,0)
\qbezier(-10,-5)(-10,0)(-18,0)
\qbezier(-10,-5)(-10,-10)(0,-10)
\qbezier(0,-10)(20,-10)(20,-5)
\qbezier(10,0)(20,0)(20,-5)
\qbezier(10,0)(0,0)(0,0)
\qbezier(10,0.2)(0,0.2)(0,0.2)
\qbezier(10,-0.2)(0,-0.2)(0,-0.2)
\qbezier(0,10)(0,0)(0,0)
\red\qbezier(-0.2,10)(-0.2,0)(-0.2,0)
\red\qbezier(0.2,10)(0.2,0)(0.2,0)
\black\qbezier[10](0,20)(0,15)(0,10)\qbezier[10](20,0)(15,0)(10,0)
\put(0,0){\blue\vector(1,1){7}}\put(9,7){\blue$\alpha$}
\black
\put(-30,0)\vtx\put(0,0)\vtx\put(10,0)\vty\put(0,10)\vty
\put(-37,-7){$a$}\put(-3,-7){$w$}\put(7,-7){$u$}\put(-7,6){$v$}
\put(-27,22){contributes $\blue\lambda x\,C_2$ to $\blue F(v)$}
\put(38,0){$\red\longleftrightarrow$}
}
\put(130,-20){
\red\qbezier(-18,0)(-30,0)(-30,0)
\qbezier(-10,-5)(-10,0)(-18,0)
\qbezier(-10,-5)(-10,-10)(0,-10)
\qbezier(0,-10)(20,-10)(20,-5)
\qbezier(10,0)(20,0)(20,-5)
\black\qbezier[20](0,20)(0,10)(0,0)\qbezier[20](20,0)(10,0)(0,0)
\put(0,0){\blue\vector(1,1){7}}\put(9,7){\blue$\alpha$}
\black
\put(-30,0)\vtx\put(0,0)\vtx\put(10,0)\vty\put(0,10)\vty
\put(-37,-7){$a$}\put(-3,-7){$w$}\put(7,-7){$u$}\put(-7,6){$v$}
\put(-27,22){contributes $\blue\lambda^2\,C_2$ to $\blue F(u)$}
}
\end{picture}
\caption{\label{fig:involution}
Involution on the Ising model configurations,
which adds or erases half-edges $vw$ and $uw$.
There are more pairs, but their relative contributions are always easy to calculate
and each pair taken together satisfies the discrete Cauchy-Riemann equations.
Note that with the chosen orientation constants $C_1$ and $C_2$ above are real.
}
\end{figure}

The strong preholomorphicity of the Ising model fermion is proved by 
constructing a bijection between 
configurations included into $F(v)$ and $F(u)$.
Indeed, erasing or adding half-edges $wu$ and $wv$ 
gives a \emph{bijection} $\omega\leftrightarrow\tilde\omega$
between configuration collections
$\brs{\omega(u)}$ and $\brs{\omega(v)}$,
as illustrated in Figure~\ref{fig:involution}.
To check \eqref{eq:shol}, it is sufficient to check
that the sum of contributions from $\omega$ and $\tilde\omega$ 
satisfies it.
Several possible configurations can be found, but essentially all
boil down to the two illustrated in Figure~\ref{fig:involution}.

Plugging the contributions from Figure~\ref{fig:involution}
into the equation \eqref{eq:shol},
we are left to check the following two identities:
\begin{equation}\lambda+\lambda\bar\lambda~=~1+\lambda \bar 1
~,~~~\lambda x+\lambda \overline{\lambda x}
~=~\lambda^2+\lambda \bar \lambda^2~.
\label{eq:condlambda}\end{equation}
The first identity always holds, while the second one is
easy to verify when $x=x_c=\sqrt{2}-1$
and $\lambda=\exp(-\pi i/4)$. 
Note that in our setup on the square lattice
$\lambda$ (or the spin $\spin$) is already fixed by the requirement that
the complex weight is well-defined, and so the second equation
in \eqref{eq:condlambda} uniquely fixes the allowed value of $x$.
In the next Section we will discuss a more general setup, allowing for
different values of the spin, corresponding to other lattice models.

To determine $F$ using its preholomorphicity,
we need to understand its behavior on the boundary.
When $z\in\partial\Omega_\epsilon$,
the \emph{winding} of the interface connecting $a$ to $z$ inside $\Omega_\epsilon$ is uniquely determined,
and coincides with the winding of the boundary itself.
This amounts to knowing $\mathrm{Arg}(F)$ on the boundary, which would be sufficient to
determine $F$ knowing the singularity at $a$
or the normalization at $b$.

In the continuous setting the condition obtained is equivalent
to the Riemann Boundary Value Problem
(a homogeneous version of the Riemann-Hilbert-Privalov BVP)
\begin{equation}
\IM\br{F(z)\cdot\br{\mathrm{tangent~to~}\partial\Omega}^{1/2}}~=~0~,
\label{eq:rbvp}\end{equation}
with the square root appearing because of the Fermionic weight.
Note that the homogeneous BVP above has conformally covariant solutions
(as  $\sqrt{dz}$-forms),
and so is well defined even in domains with fractal boundaries.
The Riemann BVP \eqref{eq:rbvp} is clearly solved by the function
$\sqrt{P_a'(z)}$, where $P$ is the Schwarz kernel  at $a$
(the complex version of the Poisson kernel),
i.e. a conformal map
$$P:\,\Omega\to\CC_+~,~~a\mapsto\infty~.$$

Showing that on the lattice $F_\epsilon$ satisfies
a discretization of the Riemann BVP \eqref{eq:rbvp}
and converges to its continuous counterpart is highly non-trivial
and a priori not guaranteed -- there exist ``logical'' discretizations
of the Boundary Value Problems, whose solutions have degenerate or no scaling limits.
We establish convergence in \cite{chelkak-smirnov-iso}
by considering the primitive $\int^z_{z_0} F^2(u)du$,
which satisfies  the Dirichlet BVP even in the discrete setting.
The big technical problem is that in the discrete case $F^2$ is no longer 
preholomorphic, so its primitive is a priori not preholomorphic
or even well-defined. 
Fortunately, in our setting the imaginary part
is still well-defined, so we can set
$$H_\epsilon(z)~:=~\frac1{2\epsilon}\IM\int^z F(z)^2dz~.$$
While the function $H$ is not exactly preharmonic,
it is approximately so, vanishes exactly on the boundary,
and is positive inside the domain.
This allows to complete the (at times quite involved) proof. 
A number of non-trivial discrete estimates is called for,
and the situation is especially difficult for general isoradial graphs.
We provide the needed tools in a separate paper \cite{chelkak-smirnov-dca}.

Though Theorem~\ref{thm:spin} establishes convergence of but one observable,
the latter (when normalized at $b$) is well behaved with respect 
to the interface traced from $a$.
So it can be used to establish the following, see \cite{chelkak-smirnov-spin}:
\begin{cor}
As mesh of the lattice tends to zero, the critical Ising interface 
in the discretization of the domain $\Omega$ with Dobrushin boundary conditions
converges to the Schramm's SLE(3) curve.
\end{cor}
Convergence is almost immediate in the topology of 
(probability measures on the space of) Loewner driving functions,
but upgrading to convergence of curves requires extra estimates, cf.
\cite{kemppainen-smirnov-curves,duminil-hongler-nolin,chelkak-smirnov-spin}.
Once interfaces are related to SLE curves,
many more properties can be established,
including values of dimensions and scaling exponents.

But even without appealing to SLE, one can use preholomorphic functions to
a stronger effect.
In a joint paper with Hongler 
\cite{hongler-smirnov-density}
we study a similar observable, when both ends of the interface are allowed to 
be inside the domain.
It turns out to be preholomorphic in both variables, except for the diagonal,
and so its scaling limit can be identified with the Green's function 
solving the Riemann BVP.
On the other hand, when two arguments are taken to be nearby, one retrieves the
probability of an edge being present in the contour representation,
or that the nearby spins are different.
This allows to establish conformal invariance of the
energy field in the scaling limit:
\begin{theorem}[Hongler, Smirnov]
\label{thm:ed}
Let $a\in\Omega$ and $\left\langle x^{\epsilon},y^{\epsilon}\right\rangle$
be the closest edge from $a\in\Omega_{\epsilon}$. Then, as $\epsilon\to0$,
we have \begin{eqnarray*}
\mathbb{E}_{+}\left[\sigma_{x}^{\epsilon}\sigma_{y}^{\epsilon}\right] & = & \frac{\sqrt{2}}{2}+\frac{l_{\Omega}\left(a\right)}{\pi}\cdot\epsilon+o\left(\epsilon\right),\\
\mathbb{E}_{\mathrm{free}}\left[\sigma_{x}^{\epsilon}\sigma_{y}^{\epsilon}\right] & = & \frac{\sqrt{2}}{2}-\frac{l_{\Omega}\left(a\right)}{\pi}\cdot\epsilon+o\left(\epsilon\right),\end{eqnarray*}
where the subscripts $+$ and $\mathrm{free}$ denote the boundary conditions
and $l_{\Omega}$ is the element of the hyperbolic metric on $\Omega$.
\end{theorem}
This confirms the
Conformal Field Theory predictions
and, as far as we know, 
for the first time provides the multiplicative constant 
in front of the hyperbolic metric.

These techniques were taken further by Hongler
in \cite{hongler-phd}, where
he showed that the (discrete) energy field in the critical Ising model 
on the square lattice
has a conformally covariant scaling limit,
which can be then identified with the corresponding 
Conformal Field Theory.
This was accomplished  by showing convergence of the discrete energy correlations
in domains with a variety of boundary conditions
to their continuous counterparts;
the resulting limits are conformally covariant
and are determined exactly.
Similar result was obtained for the scaling limit of the spin field on the domain boundary.

\section{The $O(N)$ model}\label{sec:on}

The Ising preholomorphic function was introduced
in \cite{smirnov-icm} in the setting of general $O(N)$ models on the
hexagonal lattice.
It can be further generalized to a variety of lattice models,
see the work of Cardy, Ikhlef, Rajabpour
\cite{rajabpour-cardy,ikhlef-cardy}.
Unfortunately, the observable seems only partially preholomorphic
(satisfying only some of the Cauchy-Riemann equations)
except for the Ising case.
One can make an analogy with divergence-free vector fields, which are not a priori curl-free.

\begin{figure}
\unitlength=1mm
\begin{picture}(100,75)(-20,15)
\put(-8,15){\includegraphics[width=9.24cm]{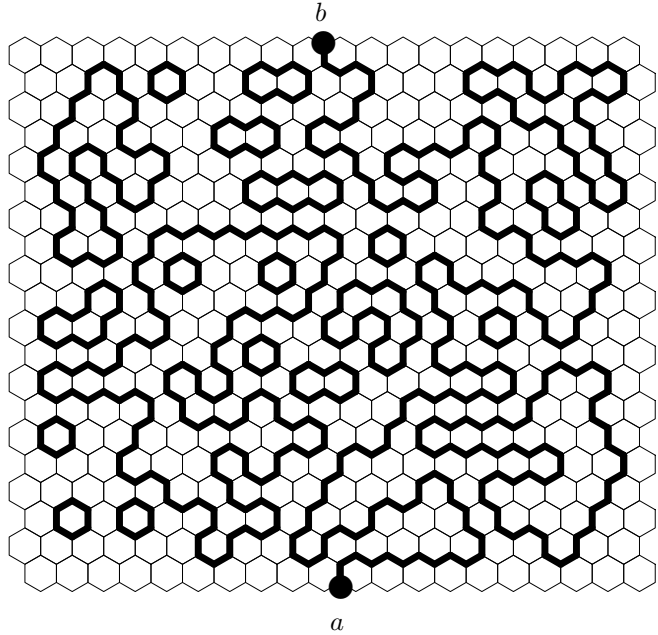}}
\put(39,13){$a$}\put(37,94){$b$}
\put(40.4,18.7){\circle*{3}}\put(38.1,91){\circle*{3}}
\end{picture}
\caption{\label{fig:on}
The high-temperature expansion of the $O({N})$ 
model leads to a gas of disjoint simple loops.
Probability of a configuration is proportional to
$N^{\mathrm{\#~loops}}~x^{\mathrm{length}}$.
We study it with Dobrushin boundary conditions:
besides loops, there is an interface between
two boundary points $a$ and $b$.
}
\end{figure}

The argument in the previous Section was adapted to the Ising case,
and some properties remain hidden behind the notion of the strong holomorphicity.
Below we present its version generalized to the $O(N)$ model,
following our joint work \cite{duminil-smirnov-saw} with Duminil-Copin.
While for $N\neq1$ we only prove
that our observable is divergence-free, 
it still turns out to be enough to deduce some global information,
establishing the Nienhuis conjecture
on the exact value of the connective constant for the hexagonal lattice:
\begin{theorem}[Duminil-Copin, Smirnov]\label{thm:saw}
On the hexagonal lattice the number $C(k)$ of distinct
simple length $k$ curves from the origin satisfies
\begin{equation}\label{eq:con}
\lim_{k\to\infty}\frac1k\log C(k)~=~\log\sqrt{2+\sqrt{2}}~.
\end{equation}
\end{theorem}

\emph{Self-avoiding walks}  on a lattice (those without self-intersections)
were proposed by chemist Flory \cite{flory-book} 
as a model for polymer chains, 
and turned out to be an interesting and extensively studied object,
see the monograph \cite{madras-slade-book}.

Using Coulomb gas formalism, physicist Nienhuis argued 
that the connective constant of the hexagonal lattice is equal to $\sqrt{2+\sqrt{2}}$,
meaning that \eqref{eq:con} holds.
He even proposed better description of the asymptotic  behavior:
\begin{equation}\label{eq:1132}
C(k)~\approx~\br{\sqrt{2+\sqrt{2}}\,}^k~k^{11/32},~~k\to\infty~.
\end{equation}
Note that while the exponential term with the connectivity constant
is lattice-dependent, the power law correction is supposed to be universal.

Our proof is partially motivated by Nienhuis' arguments, and also starts
with considering the self-avoiding walk as a special case of $O(N)$ model at $N=0$.
While a ``half-preholomorphic'' observable we construct
does not seem sufficient 
to imply conformal invariance in the scaling limit, it can be
used to establish the critical temperature,
which gives the connective constant.

The general $O(N)$ model is defined for positive integer values of $N$,
and is a generalization of the Ising model
(to which it specializes for $N=1$), with $\pm1$ spins 
replaced by points on a sphere in the $N$-dimensional space.
We work with the graphical representation, which
is obtained using the \emph{high-temperature expansion}, and makes the model
well defined for all non-negative values of $N$.

We concentrate on the hexagonal lattice in part 
because it is trivalent and so at most one contour can pass through a vertex, 
creating no ambiguities.
This simplifies the reasoning, though general graphs can also be addressed by
introducing additional weights for multiple visits of vertices.
We consider configurations $\omega$
of disjoint simple loops 
on the mesh $\epsilon$
hexagonal lattice inside domain $\Omega_\epsilon$, and two parameters:
loop-weight $N\ge0$ and (temperature-like) edge-weight $x>0$.
Partition function is then given by
$$
Z~=~\sum_{\omega}~N^{\mathrm{\#~loops}}~x^{\mathrm{length~of~contours}}~.
$$
A typical configuration is pictured in Figure~\ref{fig:on},
where we introduced Dobrushin boundary conditions:
besides loops, there is an interface $\gamma$
joining two fixed boundary points $a$ and $b$.
It was conjectured by Kager and Nienhuis \cite{kager-nienhuis}
that in the interval $N\in[0,2]$ the model has
conformally invariant scaling limits for
$x=x_{c}(N):=1/{\sqrt{2+\sqrt{2-N}}}$ and $x\in(x_c(N),+\infty)$.
The two different limits 
correspond to dilute/dense regimes,
with the interface $\gamma$ conjecturally converging to the Schramm's SLE curves
for an appropriate value of $\kappa\in[8/3,4]$ and $\kappa\in[4,8]$ correspondingly.
The scaling limit for low temperatures $x\in(0,x_c)$ is not conformally invariant.

Note that for $N=1$ we do not count the loops, thus obtaining the
low-temperature expansion of the Ising model on the dual triangular lattice.
In particular, the critical Ising corresponds to $x=1/\sqrt{3}$ by the work 
\cite{wannier-tri} of Wannier,
in agreement with Nienhuis predictions. And for $x=1$ one obtains
the critical site percolation on triangular lattice
(or equivalently the Ising model at infinite temperature).
The latter is
conformally invariant in the scaling limit by \cite{smirnov-cras,smirnov-perc}.

Note also that the Dobrushin boundary conditions make the model well-defined for $N=0$:
then we have only one interface, and no loops.
In the dilute regime this model is expected to be in the
universality class of the self-avoiding walk.

Analogously to the Ising case, we define an observable
(which is now a \emph{para\-fermion} of fractional spin)
by moving one of the ends of the interface inside the domain.
Namely, for an edge center $z$ we set
\begin{equation}\label{eq:fon}
F_\epsilon(z)~:=~\sum_{\omega(a\to z)} 
x^{\mathrm{length~of~contours}}~{\cal W}(\omega(a\to z))~,
\end{equation}
where the sum is taken over all configurations 
$\omega=\omega(a\to z)$ which have disjoint simple contours:
a number of loops and an interface $\gamma$ joining two exceptional points,
$a$ on the boundary and $z$ inside.
As before, the sum is corrected 
by a complex weight with the \emph{spin} $\spin\in\RR$:
$${\cal W}(\omega(a\to z))~:=~\exp\br{-{i}~\spin~\mathrm{winding}(\gamma, a\to z)}~,$$
equivalently we can write
$${\cal W}(\omega(a\to z))~=
~\lambda^{\# \mathrm{~signed~turns~of~}\gamma},~~\lambda:=\exp\br{- i \spin\frac\pi3}~.$$
Note that on hexagonal lattice one turn corresponds to $\pi/3$,
hence the difference in the definition of $\lambda$.

\begin{figure}
\unitlength=1.4mm
\begin{picture}(80,55)(-10,10)
\put(-8,10){\includegraphics[width=11.2cm]{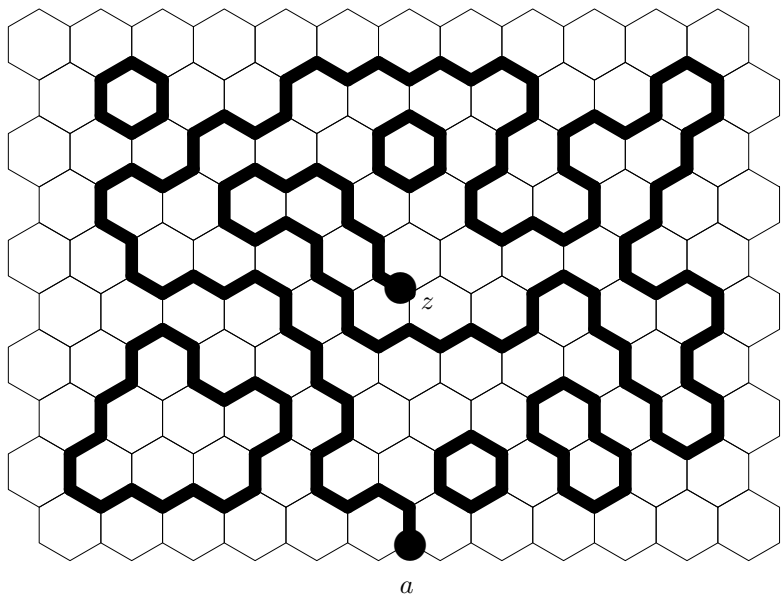}}
\put(34,10){$a$}\put(35,14.5){\circle*{3}}
\put(36,37){$z$}\put(34.1,39.0){\circle*{3}}
\end{picture}
\caption{\label{fig:onfermion}
To obtain the parafermionic observable in the $O(N)$ model
we consider configurations with an interface joining a boundary point $z$ to an 
interior point $z$ and weight them by a complex weight
depending on the winding of the interface.}
\end{figure}

Our key observation is the following
\begin{lemma}\label{lem:onrel}
For $N\in[0,2]$, set $2\cos\br{\theta}=N$
with parameter $\theta\in[0,\pi/2]$.
Then for  
\begin{eqnarray}
\label{eq:x}
\spin~=~\frac{\pi-3\theta}{4\pi}~,&
~~~~x^{-1}=~{2\cos\br{\frac{\pi+\theta}4}}
~=~\sqrt{2-\sqrt{2-N}},~~\mathrm{or}\\
\label{eq:xt}
\spin~=~\frac{\pi+3\theta}{4\pi}~,&
~~~~x^{-1}=~{2\cos\br{\frac{\pi-\theta}4}}
~=~\sqrt{2+\sqrt{2-N}}~,
\end{eqnarray}
the observable $F$ 
satisfies the following relation for every vertex $v$ inside $\Omega_\epsilon$:
\begin{equation}\label{eq:onrel}
(p-v)F(p)+(q-v)F(q)+(r-v)F(r)~=~0~,
\end{equation}
where $p$, $q$, $r$ are the mid-edges of the three edges adjacent to $v$.
\end{lemma}
Above solution \eqref{eq:x} corresponds to the dense,
and \eqref{eq:xt} -- to the dilute regime.
Note that identity \eqref{eq:onrel} is a form of the first Kirchhoff's law,
but apart from the Ising case $N=1$ we cannot verify the second one.

\begin{figure}
\def\vtx{\black\circle*{2}}
\def\vty{{\red\circle{2}}}
\unitlength=0.7mm
\begin{picture}(220,100)(-10,-40)
\thicklines
\put(30,25){
\put(-21,17){$\blue C_1$ to $\blue F(p)$}
\qbezier(-6,0)(-10,0)(-14,0)
\qbezier(-6,0)(-0.5,0)(3,5)
\qbezier(10,-10)(10,15)(3,5)
\put(10,-10)\vtx\put(6,-13){$a$}
\put(-20,0){\put(0,0)\vty
\put(6,0)\vtx\put(-3,-5.2)\vtx\put(-3,5.2)\vtx
\qbezier[12](0,0)(6,0)(12,0)
\qbezier[12](0,0)(-3,-5.2)(-6,-10.4)
\qbezier[12](0,0)(-3,5.2)(-6,10.4)
\put(1,-4){$v$}
}}
\put(85,25){
\put(-21,17){$\blue x\bar\lambda\,C_1$ to $\blue F(q)$}
\qbezier(-6,0)(-10,0)(-20,0)
\qbezier(-6,0)(-0.5,0)(3,5)
\qbezier(10,-10)(10,15)(3,5)
\put(10,-10)\vtx\put(6,-13){$a$}
\put(-20,0){\put(0,0)\vty
\put(6,0)\vtx\put(-3,-5.2)\vtx\put(-3,5.2)\vtx
\qbezier[12](0,0)(-3,-5.2)(-6,-10.4)
\qbezier[12](0,0)(-3,5.2)(-6,10.4)
\qbezier(0,0)(-3,5.2)(-3,5.2)
\put(1,-4){$v$}
}}
\put(140,25){
\put(-21,17){$\blue x\lambda\,C_1$ to $\blue F(r)$}
\qbezier(-6,0)(-10,0)(-20,0)
\qbezier(-6,0)(-0.5,0)(3,5)
\qbezier(10,-10)(10,15)(3,5)
\put(10,-10)\vtx\put(6,-13){$a$}
\put(-20,0){\put(0,0)\vty
\put(6,0)\vtx\put(-3,-5.2)\vtx\put(-3,5.2)\vtx
\qbezier[12](0,0)(-3,-5.2)(-6,-10.4)
\qbezier[12](0,0)(-3,5.2)(-6,10.4)
\qbezier(0,0)(-3,-5.2)(-3,-5.2)
\put(1,-4){$v$}
}}
\put(30,-20){
\put(-21,17){$N\,\blue C_2$ to $\blue F(p)$}
\qbezier(-6,0)(-10,0)(-14,0)
\qbezier(-6,0)(-0.5,0)(3,5)
\qbezier(10,-10)(10,15)(3,5)
\put(10,-10)\vtx\put(6,-13){$a$}
\put(-20,0){\put(0,0)\vty
\put(6,0)\vtx\put(-3,-5.2)\vtx\put(-3,5.2)\vtx
\qbezier[12](0,0)(6,0)(12,0)
\qbezier(0,0)(-3,5.2)(-6,10.4)
\qbezier(-9,5)(-9,15.6)(-6,10.4)
\qbezier(-9,5)(-9,0)(-12,-5)
\qbezier(-10,-12)(-15,-10)(-12,-5)
\qbezier(-10,-12)(-7.5,-13)(-6,-10.4)
\qbezier(0,0)(-3,-5.2)(-6,-10.4)
\put(1,-4){$v$}
}}
\put(85,-20){
\put(-21,17){$\bar\lambda^4\,\blue C_2$ to $\blue F(q)$}
\qbezier(-6,0)(-10,0)(-14,0)
\qbezier(-6,0)(-0.5,0)(3,5)
\qbezier(10,-10)(10,15)(3,5)
\put(10,-10)\vtx\put(6,-13){$a$}
\put(-20,0){\put(0,0)\vty
\put(6,0)\vtx\put(-3,-5.2)\vtx\put(-3,5.2)\vtx
\qbezier(0,0)(6,0)(12,0)
\qbezier[12](0,0)(-3,5.2)(-6,10.4)
\qbezier(-3,5.2)(-3,5.2)(-6,10.4)
\qbezier(-9,5)(-9,15.6)(-6,10.4)
\qbezier(-9,5)(-9,0)(-12,-5)
\qbezier(-10,-12)(-15,-10)(-12,-5)
\qbezier(-10,-12)(-7.5,-13)(-6,-10.4)
\qbezier(0,0)(-3,-5.2)(-6,-10.4)
\put(1,-4){$v$}
}}
\put(140,-20){
\put(-21,17){$\lambda^4\,\blue C_2$ to $\blue F(r)$}
\qbezier(-6,0)(-10,0)(-14,0)
\qbezier(-6,0)(-0.5,0)(3,5)
\qbezier(10,-10)(10,15)(3,5)
\put(10,-10)\vtx\put(6,-13){$a$}
\put(-20,0){\put(0,0)\vty
\put(6,0)\vtx\put(-3,-5.2)\vtx\put(-3,5.2)\vtx
\qbezier(0,0)(6,0)(12,0)
\qbezier[12](0,0)(-3,-5.2)(-6,-10.4)
\qbezier(0,0)(-3,5.2)(-6,10.4)
\qbezier(-9,5)(-9,15.6)(-6,10.4)
\qbezier(-9,5)(-9,0)(-12,-5)
\qbezier(-10,-12)(-15,-10)(-12,-5)
\qbezier(-10,-12)(-7.5,-13)(-6,-10.4)
\qbezier(-3,-5.2)(-3,-5.2)(-6,-10.4)
\put(1,-4){$v$}
}}
\end{picture}
\caption{\label{fig:triplets}
Configurations with the interface ending at one of the three neighbors of $v$
are grouped into triplets by adding or removing half-edges around $v$.
Two essential examples of triplets are pictured above, along with their relative 
contributions to the identity \eqref{eq:fon}.}
\end{figure}
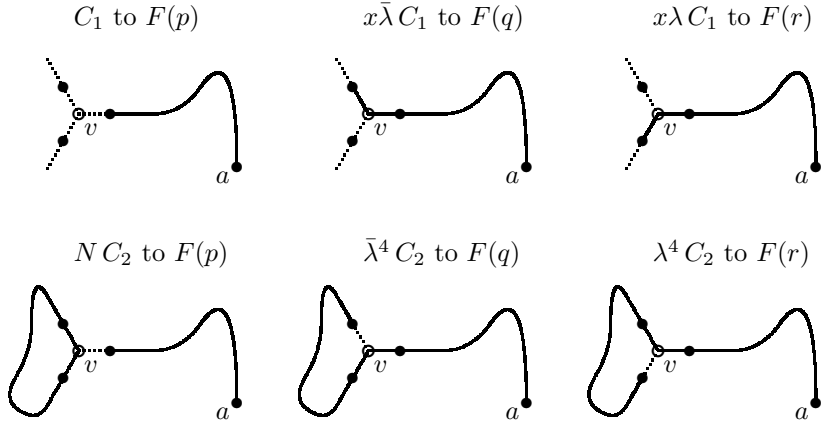

To prove Lemma~\ref{lem:onrel}, we note that
configurations with an interface arriving at $p$, $q$ or $r$
can be grouped in triplets,
so that three configurations differ only in immediate vicinity
of $v$, see Figure~\ref{fig:triplets}.
It is enough then to check that contributions
of three configurations to \eqref{eq:onrel} sum up to zero.
But the relative weights of  configurations 
in a triplet are easy to write down
as shown in Figure~\ref{fig:triplets},
and the coefficients in the identity \eqref{eq:onrel}
are proportional to the three cube roots of unity:
$1$, $\tau:=\exp(i2\pi/3)$, $\bar\tau$
(if the neighbors of $v$ are taken in the counterclockwise order).
Therefore we have to check just two identities:
\begin{eqnarray*}
N~+~\tau\,\bar\lambda^4~+~\bar\tau\,\lambda^4~=~0~,\\
1~+~\tau\,x\bar\lambda~+~\bar\tau\,x\lambda~=~0~.
\end{eqnarray*}
Recalling that  $\lambda=\exp\br{-i\spin{\pi}/{3}}$, the equations above can be recast as
\begin{eqnarray*}
-\frac{2\pi}{3}-4\spin\frac{\pi}{3}~=~\pm\br{\pi-\theta}+2\pi k~,~~~k\in\ZZ~,\\
x~=~-1\left/\br{2\cos\br{\frac{\br{2+\spin}\pi}{3}}}\right.~.
\end{eqnarray*}
The first equation implies
that
\begin{equation}\label{eq:sigma}
\spin~=~\pm\br{-\frac34+\frac{3\theta}{4\pi}}-\frac12-\frac32 k~,~~~k\in\ZZ~,
\end{equation}
and the second equation then determines the allowed value of $x$ uniquely.
Most of the solutions of \eqref{eq:sigma} lead to observables symmetric
to the two main ones, which are provided by 
solutions to the equations \eqref{eq:x} and \eqref{eq:xt}.

When we set $N=0$, there are no loops, and configurations
contain just an interface from $a$
to $z$, weighted by $x^{\mathrm{length}}$. This corresponds to taking $\theta=\pi/2$
and one of the solutions is given by $\spin=5/8$ and $x_c=1/\sqrt{2+\sqrt2}$,
as predicted by Nienhuis.
To prove his prediction, 
we observe that summing the identity \eqref{eq:onrel} 
over all interior vertices implies that
$$\sum_{z\in\partial\Omega_\epsilon}F(z) \eta(z)~=~0~,$$
where the sum taken over the centers $z$ of oriented
edges $\eta(z)$ emanating
from the discrete domain $\Omega_\epsilon$ into its exterior.
Since $F(a)=1$ by definition, we conclude
that $F$ for other boundary points sums up to $1$.
As in the Ising model, the winding on the boundary is uniquely determined,
and (for this particular critical value of $x$), one observes
that considering the real part of $F$
we can get rid of the complex weights, replacing them by explicit positive 
constants (depending on the slope of the boundary).
Thus we obtain an equation 
$$\sum_{z\in\partial\Omega_\epsilon\setminus\brs{a}}~~
\sum_{\omega(a\to z)} x_c^{\mathrm{length~of~contours}}~\asymp~1~,$$
regardless of the size of the domain $\Omega_\epsilon$.
A simple counting argument then shows that the series
$$\sum_k~C(k)~x^k~=~
\sum_{\mathrm{simple~walks~from~}a~\mathrm{inside}~\CC} x^{\mathrm{length}}~,$$
converges when $x<x_c$ and diverges when $x>x_c$, 
clearly implying the conjecture.

Note that establishing the holomorphicity of our observable in the scaling limit
would allow to relate self-avoiding walk to the Schramm's SLE with $\kappa=8/3$
and together with the work \cite{lsw-saw}
of Lawler, Schramm and Werner to establish
the more precise form \eqref{eq:1132} of the Nienhuis prediction.


\section{What's next}\label{sec:quest}

Below we present a list of open questions.
As before, we do not aim for completeness, rather
we highlight a few directions we find particularly intriguing.

\begin{question}
As was discussed, discrete complex analysis is well developed
for isoradial graphs (or rhombic lattices),
see \cite{duffin-rhombic,mercat-2001,kenyon-operators,chelkak-smirnov-dca}.
Is there a more general discrete setup
where one can get similar estimates, in
particular convergence of preholomorphic functions
to the holomorphic ones in the scaling limit?
Since not every topological quadrangulation
admits a rhombic embedding \cite{kenyon-schlenker},
can we always find another embedding 
with a sufficiently nice
version of discrete complex analysis?
Same question can be posed for triangulations,
with variations of the first definition by Isaacs \eqref{eq:cr1},
like the ones in the work of Dynnikov and Novikov \cite{dynnikov-novikov-mmj}
being promising candidates.
\end{question}

\begin{question}
Variants of the Ising observable were used by Hongler and Kyt\"ol\"a
to connect interfaces in domains with more general boundary conditions
to more advanced variants of SLE curves, see \cite{hongler-kytola}.
Can one use some version of this observable
to describe the spin Ising loop soup by a collection
of branching interfaces, which converge to a branching SLE tree 
in the scaling limit?
Similar argument os possible for the random cluster representation of the Ising model,
see \cite{kemppainen-smirnov-fk3}.
Can one construct the energy field more explicitly than in 
\cite{hongler-phd}, e.g. in the distributional sense?
Can one construct other Ising fields?
\end{question}

\begin{question}
So far ``half-preholomorphic'' parafermions similar
to ones discussed in this paper have been found in a number of models,
see \cite{smirnov-icm,riva-cardy-hol,rajabpour-cardy,ikhlef-cardy},
but they seem fully preholomorphic only in the Ising case.
Can we find the other half of the Cauchy-Riemann equations,
perhaps for some modified definition?
Note that it seems unlikely that one can establish conformal invariance
of the scaling limit operating with only half of the Cauchy-Riemann equations,
since there is no conformal structure present.
\end{question}

\begin{question}
In the case of the self-avoiding walk, an observable satisfying only a half of 
the Cauchy-Riemann
equations turned out to be enough to derive the value of the connectivity constant
\cite{duminil-smirnov-saw}.
Since similar observables are available for all other $O(N)$ models, can we use
them to establish the critical temperature values predicted by Nienhuis?
Our proof cannot be directly transfered, since some counting estimates use 
the absence of loops.
Similar question can be asked for other models.
\end{question}

\begin{question}
If we cannot establish the preholomorphicity
of our observables exactly, can we try to establish
it approximately? With appropriate estimates that would allow to obtain
holomorphic functions in the scaling limit
and hence prove conformal invariance of the models concerned.
Note that such more general approach
worked for the critical site percolation
on the triangular lattice  \cite{smirnov-cras, smirnov-perc},
though approximate preholomorphicity was a consequence of  exact identities
for quantities similar to discrete derivatives.
\end{question}

\begin{question}
Can we find other preholomorphic observables
besides ones mentioned here and in \cite{smirnov-icm}?
It is also peculiar that all the models
where preholomorphic observables were found so far
(the dimer model, the uniform spanning tree, the Ising model, percolation, etc.)
can be represented as dimer models.
Are there any models in other universality classes,
admitting a dimer representation?
Can then Kenyon's techniques \cite{kenyon-introduction,kenyon-lectures}
 be used to find preholomorphic
observables by considering the Kasteleyn's matrix and the coupling function?
\end{question}

\begin{question}
Throughout this paper we were concerned with linear discretizations
of the Cauchy-Riemann equations.
Those seem more natural in the probabilistic context,
in particular they might be easier to relate to the SLE martingales,
cf. \cite{smirnov-icm}.
However there are also well-known non-linear versions
of the Cauchy-Riemann equations.
For example, the following 
version of the Hirota equation
for a complex-valued function $F$
arises in the context of the circle packings,
see e.g. \cite{bobenko-mercat-suris}:
\begin{equation}\label{eq:hirota}
\frac{\br{F(z+i\epsilon)-F(z-\epsilon)}\br{F(z-i\epsilon)-F(z+\epsilon)}}
{\br{F(z+i\epsilon)-F(z+\epsilon)}\br{F(z-i\epsilon)-F(z-\epsilon)}}~=~-1~.
\end{equation}
Can we observe this or a similar equation in the probabilistic context
and use it to establish conformal invariance of some model?
Note that plugging into the equation \eqref{eq:hirota} 
a smooth function,
we conclude that to satisfy it approximately
it must obey the identity
$$\br{\partial_x F(z)}^2+\br{\partial_y F(z)}^2~=~0~.$$
So in the scaling limit \eqref{eq:hirota} can be factored into
the Cauchy-Riemann equations and their complex conjugate,
thus being in some sense linear.
It does not seem possible to obtain ``essential'' non-linearity using just four points,
but using five points one can create one,
as in the next question.
\end{question}

\begin{question}
A number of non-linear identities was discovered for
the correlation functions in the Ising model,
starting with the work of Groeneveld, Boel and Kasteleyn
\cite{groeneveld-boel-kasteleyn,boel-kasteleyn}.
We do not want to analyze the extensive literature to-date,
but rather pose a question: 
can any of these relations be used to define
discrete complex structures and pass to the scaling limit?
In two of the early papers by McCoy, Wu and Perk
\cite{mccoy-wu-nonlin,perk-nonlin},
a quadratic difference relation was observed in the full plane Ising model
first on the square lattice, and then on a general graph.
To better adapt to our setup, we rephrase this relation
for the correlation $C(z)$ of two spins (one at the origin and another at $z$)
in the Ising model at criticality on the mesh $\epsilon$ square lattice.
In the full plane, one has 
\begin{equation}\label{eq:nonlin}
C(z+i\epsilon)C(z-i\epsilon)+C(z+\epsilon)C(z-\epsilon)=2C(z)^2~.
\end{equation}
Note that $C$ is a \emph{real-valued} function,
and the equation \eqref{eq:nonlin} is a discrete form of
the identity
$$C(z)\Delta C(z)+\abs{\nabla C(z)}^2=0~.$$
The latter is conformally invariant, and is solved by moduli
of analytic functions.
Can one write an analogous to \eqref{eq:nonlin} identity
in domains with boundary, perhaps approximately?
Can one deduce conformally invariant scaling limit
of the spin correlations in that way?
\end{question}

\begin{question}
Recently there was a surge of interest in random planar graphs and their 
scaling limits, see e.g. \cite{ds-prl,MR2438999}.
Can one find observables on random planar graphs
(weighted by the partition function of some lattice model)
which after an appropriate embedding
(e.g. via a circle packing or  a piecewise-linear Riemann surface)
are preholomorphic?
This would help to show that planar maps converge
to the Liouville Quantum Gravity in the scaling limit.
\end{question}

\begin{question}
Approach to the two-dimensional
integrable models described here is
in several aspects similar to the older
approaches based on the Yang-Baxter relations \cite{baxter-book}.
Some similarities are discussed in Cardy's paper \cite{cardy-smm}.
Can one find a direct link between the two approaches?
It would also be interesting to find a link
to the three-dimensional consistency relations
as discussed in \cite{bazhanov}.
\end{question}

\begin{question}
Recently Kenyon investigated the Laplacian on the vector bundles over graphs
in relation to the spanning trees \cite{kenyon-bundle}.
Similar setup seems natural for the Ising observable we discuss.
Can one obtain more information about the Ising and other models
by studying difference operators on vector bundles
over the corresponding graphs?
\end{question}

\begin{question}
Can anything similar be done for the three-dimensional models?
While preholomorphic functions do not exist here,
preharmonic vector fields are well-defined and appear naturally
for the Uniform Spanning Tree and the Loop Erased Random Walk.
To what extent can they be used?
Can one find any other difference equations
in three-dimensional lattice models?
\end{question}

\bibliographystyle{alpha}

\end{document}